\documentclass[bibyear]{aa}  
\usepackage{graphicx}
\usepackage[]{natbib}       
\usepackage{txfonts}
\usepackage{graphicx}	
\usepackage{amsmath}	
\usepackage{amssymb}
\usepackage{upgreek}
\usepackage{ulem}
\usepackage{xcolor}
\newcommand{\msol}{\mathrm{M_{\odot}}}
\newcommand{\kgmcube}{\mathrm{kg\,m^{-3}}}
\newcommand{\kgmsq}{\mathrm{kg\,m^{-2}}}
\newcommand{\kgssquare}{\mathrm{kg.s^{-2}}}
\newcommand{\jmsqare}{\mathrm{J\,m^{-2}}}

\newcommand{\msec}{\mathrm{m\,s^{-1}}}
\newcommand{\au}{\mathrm{au}}
\newcommand\tenpow[1]{10^{#1}}
\newcommand\xtenpow[1]{\ensuremath{{\:\times\:}10^{#1}}}

\defcitealias{Rozner_2020_Erosiona}{R20}
\defcitealias{Grishin_2020_Erosionb}{G20}

\begin{document} 

   \title{Aeolian erosion in protoplanetary discs:\\ How impactful it is on dust evolution?}

   \subtitle{}

   \author{Stéphane Michoulier
          \inst{1}
          \and
          Jean-François Gonzalez\inst{1}
          \and
          Evgeni Grishin\inst{2}
          \and
          Clément Petetin\inst{1}
          }

   \institute{Universite Claude Bernard Lyon 1, CRAL UMR5574, ENS de Lyon, CNRS, Villeurbanne, F-69622, France\\
              \email{jean-francois.gonzalez@ens-lyon.fr}
        \and
             Monash Centre for Astrophysics (MoCA) and School of Physics and Astronomy, Monash University, Vic. 3800, Australia
             }

   \date{Received 13 November 2023; accepted 26 February 2024}

 
  \abstract
   { Many barriers prevent dust to form planetesimals via coagulation in protoplanetary discs, such as bouncing, collisional fragmentation or aeolian erosion. Modelling dust and the different phenomena that can alter its evolution is therefore needed. Multiple solutions have been proposed, but still need to be confirmed.}
   { In this paper, we explore the role aeolian erosion plays in the evolution of dust.}
   { We use a monodisperse model to account for dust growth and fragmentation, implemented in a 1D model to compute the evolution of single grains and a 3D SPH code to compute the global evolution of dust and gas. We test the erosion model in our code and ensured it matches previous results.}
   { With a model of disc reproducing observations, we show with both 1D and 3D studies that erosion is not significant during the evolution of dust when we take fragmentation into consideration. With a low-viscosity disc, fragmentation is less of a problem, but grain growth is also less important, preventing the formation of large objects anyway. In dust traps, close to the star, erosion is also not impactful, even when fragmentation is turned off.}
   { We show in this paper that aeolian erosion is negligible when radial drift, fragmentation and dust traps are taken into account and does not alter the dust evolution in the disc. However, it can have an impact on later stages, i.e. when the streaming instability forms large clumps close to the star, or when planetesimals are captured.}

   \maketitle
%

\section{Introduction}

Protoplanetary discs consist of gas and dust that orbit young stars and provide the necessary material for the agglomeration and growth of planetesimals, the building blocks of planets. The dynamical and physical processes occurring within these discs play a crucial role in shaping the characteristics and composition of planetary systems. Among these processes, aeolian erosion of large dust particles might influence the dynamics and evolution of the dust in the inner regions of protoplanetary discs \citep{Blum_2000,Wurm_2001}.

Aeolian erosion is a process where dust particles are ejected from a larger object due to the combined action of gas drag and turbulent motions within the disc. It has been studied by \citet{Blum_2000,Wurm_2001,Paraskov_2006_erosion}, and more recently by \citet{Rozner_2020_Erosiona} and \citet{Grishin_2020_Erosionb} (hereafter \citetalias{Rozner_2020_Erosiona} and \citetalias{Grishin_2020_Erosionb}). They showed that large aggregates can be eroded in a short time, typically ranging from a few years to a few thousand years, with sizes going from several hundred metres down to a couple of centimetres \citepalias{Rozner_2020_Erosiona,Grishin_2020_Erosionb}. 
This process is therefore believed to destroy large boulders very efficiently and impact the evolution of grains in the inner region of the disc. Hence, erosion is another barrier to dust growth from small sizes to kilometric objects.

In addition to aeolian erosion, fragmentation of dust particles is a process that can destroy grains in the inner regions of protoplanetary discs. When dust particles collide at high velocities, they may experience catastrophic disruptions, leading to the so-called fragmentation barrier \citep{weidenschilling_formation_1993,dominik_physics_1997,blum_growth_2008}. 
Fragmentation thresholds or material properties to model collisions are still not fully understood, with many uncertainties remaining.
Moreover, dust experiences radial drift during its growth in the disc due to gas drag. Grains of a few centimetres to metres drift very efficiently, due to marginal coupling to the gas, and are accreted onto the star \citep{whipple_certain_1972}, which defines the radial drift barrier \citep{weidenschilling_aerodynamics_1977}.
In order to prevent radial drift and the loss of material onto the star, several solutions have been proposed in order to help the formation of planetesimals. Some rely on the capture of dust in pressure maxima, others bypass the barriers to dust growth. For instance, vortices have been explored \citep{barge_did_1995,meheut_formation_2012,loren-aguilar_toroidal_2015} to trap dust and form clumps directly from gravitational collapse. Snow lines \citep{kretke_grain_2007,brauer_planetesimal_2008,drazkowska_modeling_2014,vericel_self-induced_2020} and self-induced dust traps \citep{gonzalez_self-induced_2017,vericel_self-induced_2020,vericel_dust_2021} form local pressure maxima, stopping the radial drift and helping grain growth by the increase of the local dust density. Other properties of dust have also been investigated like grain porosity \citep{ormel_dust_2007,suyama_numerical_2008,okuzumi_numerical_2009,okuzumi_rapid_2012,kataoka_fluffy_2013,garcia_evolution_2020}, allowing grains to grow faster and to larger sizes, while being less sensitive to fragmentation.
Additionally, other processes related to instabilities have been under investigation these recent years, mostly with the streaming instability \citep{youdin_streaming_2005,youdin_protoplanetary_2007,yang_concentrating_2017,schafer_initial_2017,auffinger_linear_2018,li_demographics_2019} which allows dust to directly form boulders from pebbles, bypassing the different barriers grains might undergo during their evolution.
\citet{Grishin_2019} looked at a different solution where planetesimals can be captured in protoplanetary discs early. This had also been shown to occur in earlier stages, such as molecular clouds \citep{Pfalziner_2021a,Pfalziner_2021b}.

In this paper, we study the importance of aeolian erosion in relation to fragmentation and its importance in disc inner regions. 
In a previous paper, we compared the importance of fragmentation and rotational disruption of porous grains \citep{Michoulier_Gonzalez_Disruption}, a new barrier introduced by \citet{tatsuuma_rotational_2021}. In this paper, however, we will not take into account porosity as the current model for aeolian erosion of \citetalias{Rozner_2020_Erosiona} and \citetalias{Grishin_2020_Erosionb} does not take into account the evolution of density. We will therefore limit ourselves to the simpler compact grains formalism.

The paper is built as follows: we first describe our dust growth and fragmentation model and introduce the model to take aeolian erosion into account in Sect.~\ref{Sc:Methods}. We then show the tests performed to make sure the implementation of the erosion module is correctly done in Sect.~\ref{Sc:Tests}. Then, we discuss the results in Sect.~\ref{Sc:Results} using the 1D code \textsc{Pamdeas}\footnote{https://github.com/StephaneMichoulier/Pamdeas.git} \citep{Michoulier_Gonzalez_Disruption} and the 3D code \textsc{Phantom} by \citet{price_phantom_2018} to show the unimportance of aeolian erosion with respect to fragmentation. Finally, we discuss our results about erosion and the limitations of the codes in Sect.~\ref{Sc:Discussion}, and end with a conclusion in Sect.~\ref{Sc:Conclusions}.

\section{Methods}\label{Sc:Methods}
To determine the importance of different barriers, we describe in the following how growth, fragmentation and erosion are taken into account in \textsc{Pamdeas}. When using \textsc{Phantom} \citep{price_phantom_2018}, all the details about the implementation of growth and fragmentation are presented in \citet{vericel_dust_2021}. The erosion model implemented in $\textsc{Pamdeas}$ and $\textsc{Phantom}$ is the same, and the growth and fragmentation models are identical, but the implementation differs slightly and are code-related.

\subsection{Dust grain growth model}\label{Ssc:Dust_grain_growth_model}
In order to model dust growth, we consider a locally uniform mass distribution of grains where collisions occur exclusively between grains of identical mass. The equation describing the size variation is given by \citet{stepinski_global_1997}:
\begin{equation}\label{Eq:Growth_dmdt_Stepinski_ter}
    \left(\frac{\mathrm{d} s}{\mathrm{d} t}\right)_\mathrm{grow}= \frac{\rho_\mathrm{d}}{\rho_\mathrm{s}} v_\mathrm{rel}.
\end{equation}
where $\rho_\mathrm{d}$ is the dust local density, $\rho_\mathrm{s}$ the grain intrinsic density and $v_\mathrm{rel}$, relative velocity during collision, is:
\begin{equation}\label{Eq:Vrel}
    v_\mathrm{rel} = \sqrt{2^{3/2}\mathrm{Ro}\,\alpha}\,c_\mathrm{g}\frac{\sqrt{\mathrm{St}}}{1+\mathrm{St}}.
\end{equation}
$\alpha$ is the turbulent viscosity parameter \citep{shakura_black_1973} and $c_\mathrm{g}$ is the gas sound speed. St is the Stokes number usually defined as the dimensionless stopping time, $\Omega_\mathrm{K}t_\mathrm{s}$, where $\Omega_\mathrm{K}$ is the Keplerian frequency and $t_\mathrm{s}$ the drag stopping time. St quantifies the coupling between gas and dust, which depends on the drag regimes and the grain size (St~$\propto s$ in the Epstein regime). $\mathrm{Ro}$, the Rossby number, is considered to be a constant equal to 3 \citep{stepinski_global_1997}. Full details of the model and its derivation are presented in \citet{laibe_growth_2008,gonzalez_accumulation_2015,gonzalez_self-induced_2017} or \citet{vericel_dust_2021}.
Naturally, the growth rate increases as the relative velocity increases, as this enhances the probability of collision occurrence. Moreover, the growth rate exhibits a linear dependence on $\rho_\mathrm{d}$, indicating that the settling of dust grains or their accumulation in dust traps favour dust growth.

%

\subsection{Fragmentation}\label{Ssc:Fragmentation}

When the relative velocity between grains exceeds a critical threshold known as $v_\mathrm{frag}$, instead of growing, the grains undergo fragmentation, as observed by \citet{Tanaka_1996}. The impact's kinetic energy becomes too high for the grain's structure to absorb, causing the bonds between the constituent monomers of the aggregate to break apart.
To quantify the mass variation during fragmentation, we can adopt a formulation similar to that proposed by \citet{stepinski_global_1997} for growth. Using a realistic approach developed by \citet{Kobayashi_fragmentation_2010} and \citet{garcia_evolution_2018} and further used by \citet{vericel_dust_2021}, the model introduces a concept of progressive or ``soft'' fragmentation, taking into account the relative velocity ($v_\mathrm{rel}$) compared to $v_\mathrm{frag}$:
\begin{equation}\label{Eq:Frag_dmdt_Garcia}
    \left(\frac{\mathrm{d} s}{\mathrm{d} t}\right)_\mathrm{frag}= - \frac{v_\mathrm{rel}^2}{v_\mathrm{rel}^2 + v_\mathrm{frag}^2}\frac{\rho_\mathrm{d}}{\rho_\mathrm{s}} v_\mathrm{rel}.
\end{equation}
When the relative velocity is near the fragmentation threshold, the mass loss becomes less pronounced. Therefore, a fragmenting grain loses approximately half of its initial mass after a collision time\footnote{It is worth noting that our definition of fragmentation aligns with the point at which half of the mass is lost. However, it’s important to acknowledge that other definitions exist. For example, \citet{Ringl_Bringa_2012,Gunkelmann_Ringl_2016} define fragmentation as the moment when at least one monomer is ejected from the main aggregate. We refer to the loss of grain mass when it’s less than half of the initial mass as erosion.} when $v_\mathrm{rel}=v_\mathrm{frag}$ or more when $v_\mathrm{rel}>v_\mathrm{frag}$. When $v_\mathrm{rel} \gg v_\mathrm{frag}$, the situation described by \citet{gonzalez_accumulation_2015} is recovered, where the entire grain fragments, losing most of its mass within a collision time.

\begin{figure*}
    \centering
    \includegraphics[width=0.48\textwidth,clip]{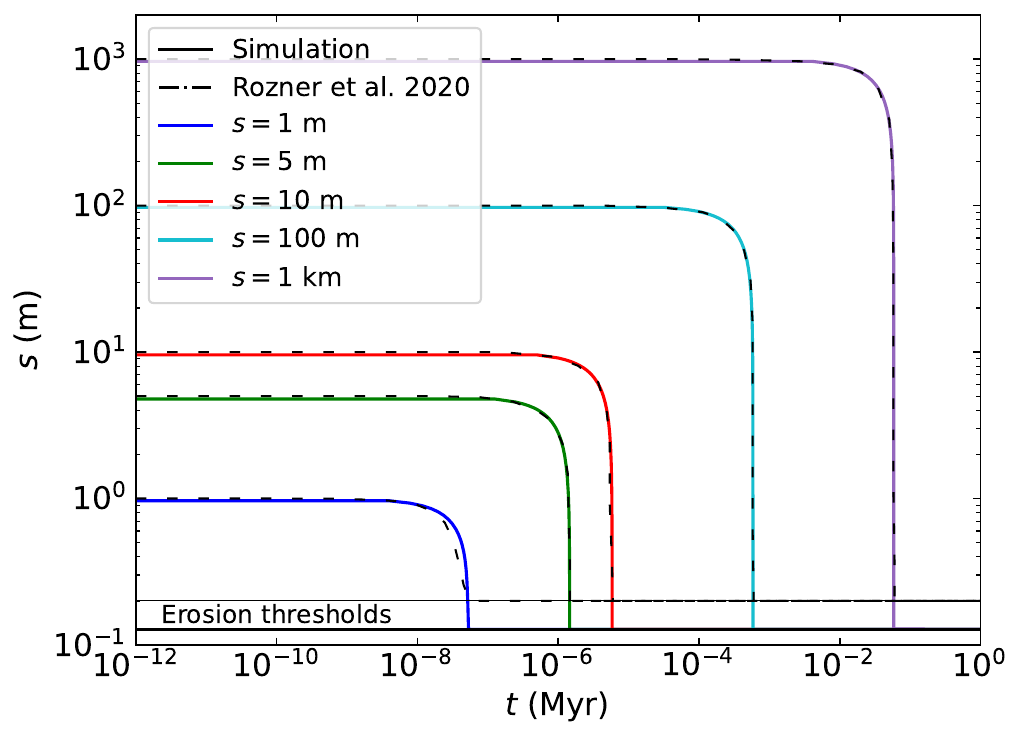}
    \includegraphics[width=0.48\textwidth,clip]{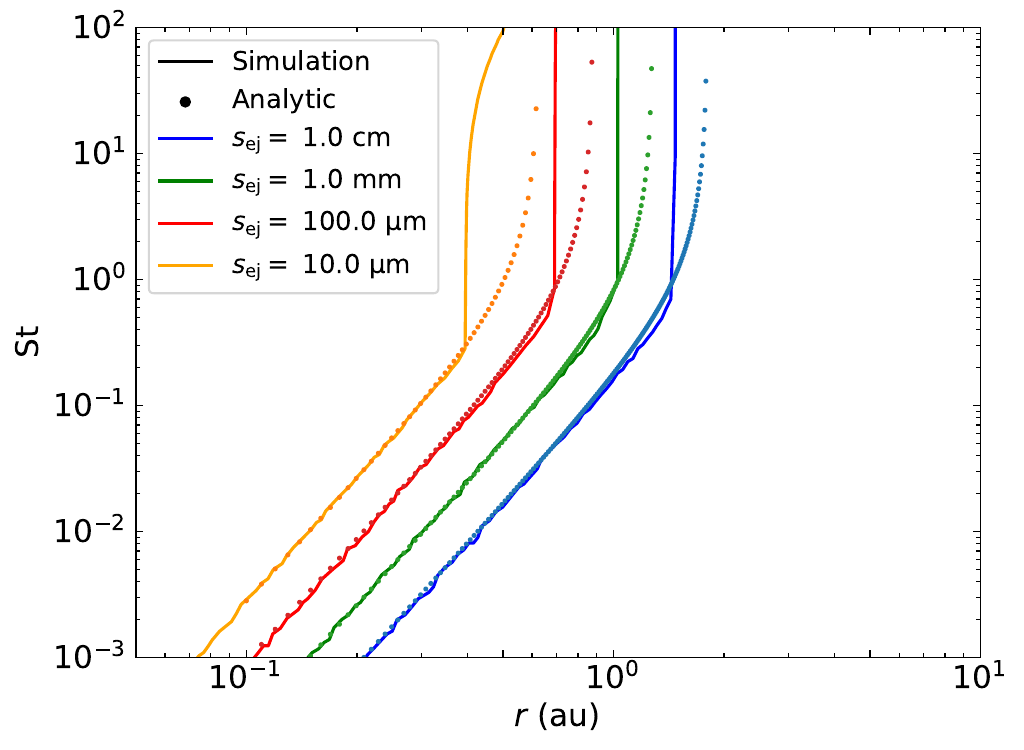}
    \caption{\textbf{Left}: Time evolution of the size $s$ of different aggregates as they are eroded. The vertical turnover in the profile gives the characteristic erosion time. The aggregates are kept at a fixed distance of $1~\au$ from the star, and the ejected dust has a size of 100 $\mu$m. Dashed lines represent data from \citetalias{Rozner_2020_Erosiona}, which are in excellent agreement with our results. \textbf{Right}: Stokes number at the erosion threshold for aggregates drifting towards the star and four different sizes of ejected grains. Dotted lines represent the analytical solution presented in \citetalias{Grishin_2020_Erosionb}.}
    \label{Fig:Erosion_test}
\end{figure*}
%
\subsection{Aeolian erosion}\label{Ssc:Aeolian_erosion}
\citetalias{Rozner_2020_Erosiona} and \citetalias{Grishin_2020_Erosionb} considered aeolian erosion by gas as a mechanism to reduce the size of large aggregates when they are decoupled from the gas. The following section is inspired by those works. During erosion, grains of size $s_\mathrm{ej}$ are ejected from the original aggregate of size $s$ ranging from one metre to one kilometre. 
If we consider a surrounding gas density $\rho_\mathrm{g}$ and a velocity difference between dust and gas $\Delta v$, 
the characteristic time for a grain to be ejected is given by:
\begin{equation}
    t_\mathrm{ej} = \frac{\Delta v}{a_\mathrm{coh}},
\end{equation}
where $a_\mathrm{coh}$ is the cohesive acceleration required for the grain to remain attached.
The work exerted by the gas on the aggregate and the energy loss due to erosion are given by \citetalias{Rozner_2020_Erosiona}:
\begin{align}
    W &= p\Delta v A t_\mathrm{ej},\\
    \Delta E &= - \Delta m \frac{\Delta v^2}{2},
\end{align}
where $p=\frac{1}{2}\rho_\mathrm{g}\Delta v^2$ is the dynamic pressure and $A$ is the effective shear surface. 
Since work and energy loss are equal, the mass loss is then:
\begin{equation}
    \Delta m = - A \rho_\mathrm{g} \frac{\Delta v^2}{a_\mathrm{coh}}.
\end{equation} 
In the case of an infinitesimal time interval $\mathrm{d} t < t_\mathrm{ej}$, $A \approx s\Delta v \mathrm{d} t$, the mass loss becomes
\begin{equation}\label{Eq:dmdteros_grishin}
    \left(\frac{\mathrm{d} m}{\mathrm{d} t}\right)_\mathrm{eros} = -\rho_\mathrm{g} \frac{\Delta v^3}{a_\mathrm{coh}}s.
\end{equation}
Currently, $a_\mathrm{coh}$ is an unknown but \citet{Shao_2000_erosion} showed that the cohesive force could be related to $a_\mathrm{coh}$.
\begin{equation}
    F_\mathrm{coh} = m_\mathrm{ej} a_\mathrm{coh} = \beta_\mathrm{eros} s_\mathrm{ej},
\end{equation}
where $m_\mathrm{ej}$ is the mass of an ejected grain and $\beta_\mathrm{eros}$ is a parameter to define.
The experimental measurements of $\beta_\mathrm{eros}$ by \citet{Heim_1999_friction_forces} and \citet{Paraskov_2006_erosion} show a strong cohesive force, resulting in a value of $\beta_\mathrm{eros} = 0.1\ \kgssquare$. With this relation between $a_\mathrm{coh}$ and $\beta_\mathrm{eros}$, we can finally rewrite Eq.~(\ref{Eq:dmdteros_grishin}) in terms of size loss as:
\begin{equation}\label{Eq:dmdteros_grishinbis}
    \left(\frac{\mathrm{d} s}{\mathrm{d} t}\right)_\mathrm{eros} = \frac{-\rho_\mathrm{g}}{3} \frac{s^2_\mathrm{ej}}{s} \frac{\Delta v^3}{\beta_\mathrm{eros}}.
\end{equation}
This equation will be used in the simulations to determine the importance of erosion.
However, for erosion to occur, the erosion threshold must first be reached. This threshold was determined by \citet{Shao_2000_erosion} and is expressed as:
\begin{equation}\label{Eq:deltav_eros_grishin}
    \Delta v_\mathrm{eros} = \sqrt{A_\mathrm{N}\frac{\gamma_\mathrm{s}}{\rho_\mathrm{g} s_\mathrm{ej}}},
\end{equation}
where $A_\mathrm{N} = 0.0123$ is a numerical constant and $\gamma_\mathrm{s} = 1.65\xtenpow{-4}~\jmsqare$, the surface energy, is the same as in \citetalias{Rozner_2020_Erosiona}. $\Delta v_\mathrm{eros}$ thus depends essentially on $s_\mathrm{ej}$.

In this work, $\Delta v = v_{\mathrm{d}} - v_{\mathrm{g}}$, the velocity difference between dust and gas is given by:
\begin{align}
    \label{Eq:Deltavr}
    \Delta v_\mathrm{r} &= \frac{(1+\varepsilon)\mathrm{St}}{(1+\varepsilon)^2+\mathrm{St}^2}v_\mathrm{drift} - \frac{\mathrm{St}^2}{(1+\varepsilon)^2+\mathrm{St}^2}v_\mathrm{visc},\\
    \Delta v_\mathrm{\theta} &= - \frac{\mathrm{St}^2}{\left(1+\epsilon \right)^2+\mathrm{St}^2} \frac{v_\mathrm{drift}}{2} - \frac{\left(1+\varepsilon\right)\mathrm{St}}{\left(1+\varepsilon \right)^2+\mathrm{St}^2} \frac{v_\mathrm{visc}}{2},
    \label{Eq:Deltavphi}
\end{align}
where $\varepsilon$ is the dust-to-gas ratio, $v_\mathrm{drift}$ is the radial drift velocity associated to the gas pressure gradient, and $v_\mathrm{visc}$ is the one due to gas motion caused by viscosity \citep[full details on the derivation can be found in][]{Michoulier_Gonzalez_Disruption}.  
$\Delta v$ is hence given by $\Delta v = \sqrt{\Delta v_\mathrm{r}^2 + \Delta v_\mathrm{\theta}^2}$. and is compared to the erosion threshold $\Delta v_\mathrm{eros}$.


%
\section{Tests}\label{Sc:Tests}
To ensure that our implementation of the equations describing erosion works as intended, we conduct tests to compare our results with those of \citetalias{Rozner_2020_Erosiona} and \citetalias{Grishin_2020_Erosionb}, as shown in Fig.~\ref{Fig:Erosion_test}.
To conduct these tests, we use the same disc parameters, which correspond to a Minimum Mass Solar Nebula model \citep{Peret_2011,Grishin_2015}.
The surface density is $\Sigma_\mathrm{g}=2\xtenpow{4}\,(r/r_0)^{-1.5}$~$\kgmsq$, and the temperature \mbox{$T_\mathrm{g}=120\,(r/r_0)^{-3/7}$~K}, with $r_0=1~\au$. This gives an aspect ratio $H/r=0.022\,(r/r_0)^{2/7}$ and a gas density \mbox{$\rho_\mathrm{g} \approx 3\xtenpow{-6}\,(r/r_0)^{-16/7}$~$\kgmcube$}. The dust-to-gas ratio is small and does not play a role here.
Figure~\ref{Fig:Erosion_test} on the left demonstrates that our simulations using \textsc{Pamdeas} perfectly reproduce the data from Figure 4 of \citetalias{Rozner_2020_Erosiona}. The characteristic erosion timescale for each aggregate size is accurately reproduced. When $\Delta v_\mathrm{eros} = \Delta v$, as the distance is kept fixed, the erosion threshold corresponds to only one value of St and hence one size. It should be noted that the size at which the erosion threshold is reached is slightly different (127 cm instead of 200 cm in \citetalias{Rozner_2020_Erosiona}). This difference likely arises from the way $\Delta v$ is computed, which differs between this work and \citetalias{Rozner_2020_Erosiona} and \citetalias{Grishin_2020_Erosionb}. Although the physical size is different, what determines the relative velocity is the Stokes number, which is the same, as shown in the right panel.

\begin{figure*}
    \centering
    \includegraphics[width=0.48\textwidth,trim=.05cm .05cm .05cm .05cm,clip]{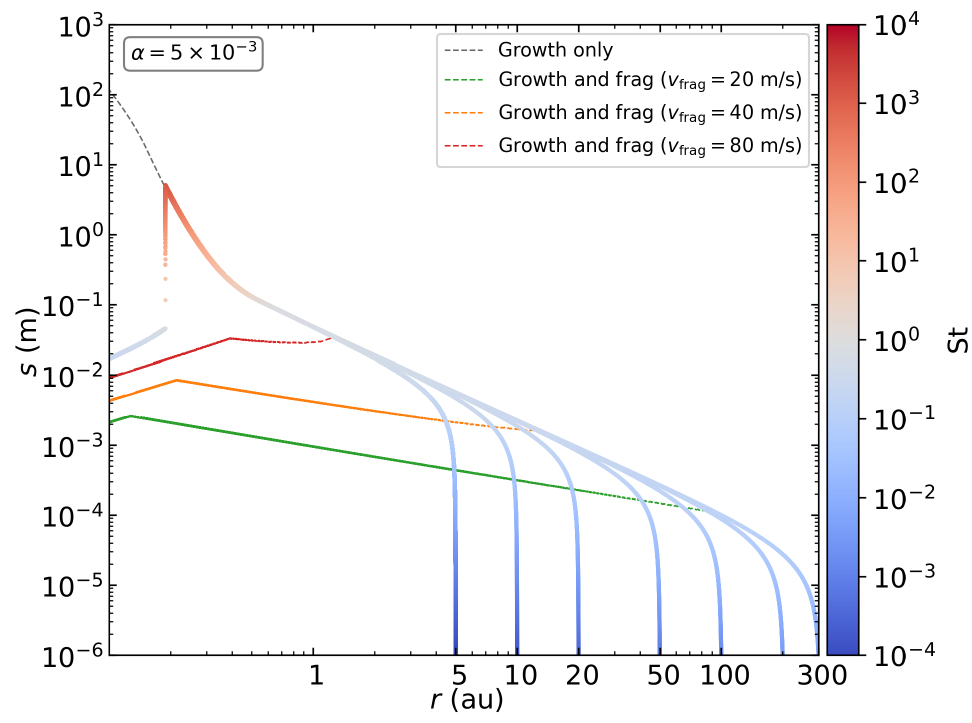}
    \includegraphics[width=0.48\textwidth,trim=.05cm .05cm .05cm .05cm,clip]{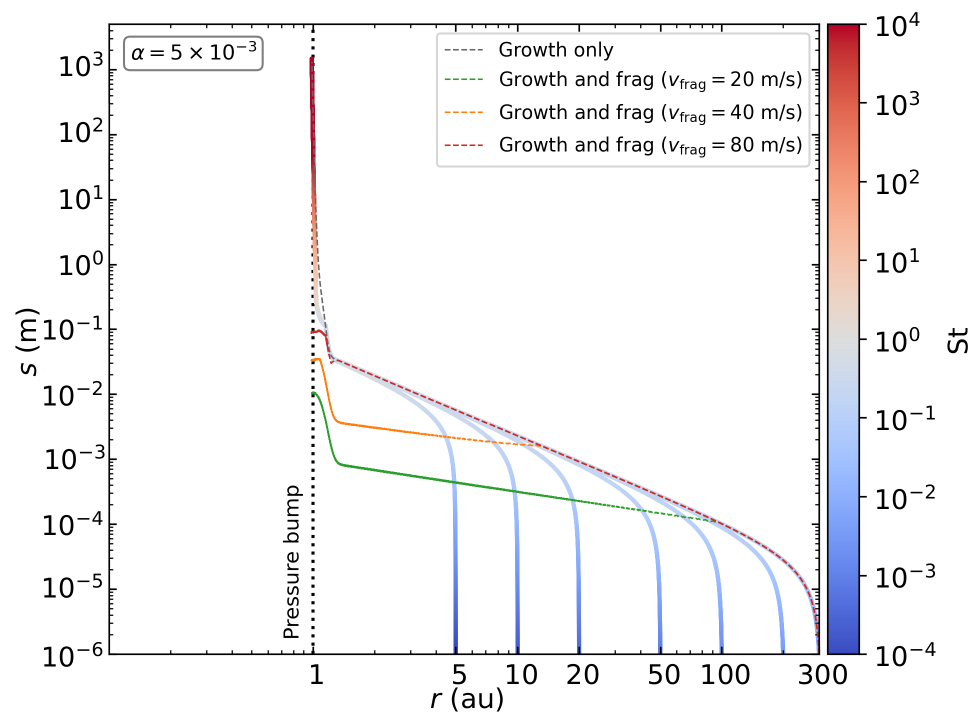}
    \caption{\textbf{Left}: Evolution of the size $s$ of grains experiencing growth and erosion only, as they grow and drift inwards from different initial distances $r$ from the star. The colour represents the Stokes number. The gray dashed line represents the evolution without erosion, and the green, orange and red lines represent the evolution with growth and fragmentation using a fragmentation threshold of 20, 40 and 80 $\msec$. The size of the ejected grains is $s_\mathrm{ej}=1$ mm, a typical value used by \citetalias{Rozner_2020_Erosiona} and \citetalias{Grishin_2020_Erosionb}. \textbf{Right}: Same as the left panel, but with a pressure maximum that traps dust at $1~\au$.}
    \label{Fig:Erosion_growth-3}
\end{figure*}

On the right, Fig.~\ref{Fig:Erosion_test} shows the critical Stokes number ($\mathrm{St}$) at the erosion threshold. In this case, the grains to be eroded have an initial size of 10 metres, and are initially placed in the outer disc. Grains then drift toward the star until they cross the erosion threshold. The data are compared with the analytical solution that give the erosion threshold in terms of critical Stokes number, a polynomial of degree 5, which takes into account both laminar and turbulent gas flow \citepalias[see Eq.~(8) in][]{Grishin_2020_Erosionb} around the aggregate. To achieve this, one writes the sum of $\Delta v$ for laminar and turbulent gas flow as a function of $\mathrm{St}$, which equals $\Delta v_\mathrm{eros}$ at the erosion threshold, and then numerically solves for $\mathrm{St}$.
As the size of the ejected grains increase, erosion becomes easier as erosion starts at larger distances $r$ compared to smaller $s_\mathrm{ej}$. The simulations also show excellent agreement with the analytical solution. The difference occurs at large $\mathrm{St}$ and is due to the fact that a grain evolving in \textsc{Pamdeas} takes a non-zero time to be eroded and reach the equilibrium.

\section{Results}\label{Sc:Results}
\subsection{Setup}
To be closer to reality than the Minimum Mass Solar Nebula (MMSN) model, we use a disc model that reproduces observations, presented in \citet{williams_parametric_2014}. For \textsc{Pamdeas}, the masses of the star and of the disc are set to $M_* = 1~\msol$ and $M_\mathrm{disc} = 0.01~\msol$. The inner and outer radii are $R_\mathrm{in}=0.1~\au$ and $R_\mathrm{out}=300~\au$, while the disc aspect ratio is $H/R_0=0.0895$ with a reference radius $R_0=100~\au$.
The density and temperature profiles exponents are $p=1$ and $q=0.5$. The turbulent viscosity parameter \citep{shakura_black_1973} is set to $\alpha = 5\times 10^{-3}-5\times 10^{-4}$. A typical dust-to-gas ratio of 1\% is used. We use silicates with $\rho_\mathrm{s}= 2.7~\kgmcube$ and initial size $s_0=1~\mu$m, with different fragmentation thresholds $v_\mathrm{frag} = 20-40-80~\msec$. For erosion, we use $\beta_\mathrm{eros} = 0.1\ \kgssquare$, and the ejected grain size $s_\mathrm{ej} = 1$ mm, the largest physical value to maximise the effect of erosion \citepalias{Rozner_2020_Erosiona,Grishin_2020_Erosionb}, as smaller $s_\mathrm{ej}$ would reduce $\left(\frac{\mathrm{d} s}{\mathrm{d} t}\right)_\mathrm{eros}\propto s_\mathrm{ej}^2$. In addition, one of the assumptions in the erosion model is that the cohesive force $F_\mathrm{coh}$ is proportional to the grain size, which, according to \citetalias{Grishin_2020_Erosionb} is valid up to 1 mm. Hence, we will only use this value in this paper. To model a pressure bump centred at $1~\au$, we simply add to the initial profile a Gaussian with a width of $0.1~\au$ and a height of 20 times the initial surface at $1~\au$. The local dust-to-gas ratio in the pressure bump is set to 1, a typical value expected in dust traps where dust is settled and concentrated.
As 3D simulations are more expensive, we only model the inner disc with \textsc{Phantom}, keeping $M_*$ and $R_\mathrm{in}$ the same but taking $R_\mathrm{out}=5~\au$ and $R_0=1~\au$, which gives $M_\mathrm{disc} = 1.64\xtenpow{-4}~\msol$ in order to have the same surface density, with $\alpha = 5\times 10^{-3}$. We use for all 3D simulations $\tenpow{6}$ particles. Simulations start with gas only, in order to prevent artefacts due to gas relaxation. After 10 orbits at 5 $\au$ ($\sim 120$ yr), the disc is relaxed, and dust is added with a uniform dust-to-gas ratio of 1\%, and the grain size is initialized at $s= 100~\mu$m, with $v_\mathrm{frag} = 80~\msec$.
The other parameters remain the same.

\subsection{1D study}\label{Ssc:1D_Study}
In \citetalias{Rozner_2020_Erosiona} and \citetalias{Grishin_2020_Erosionb}, erosion near the star is presented for certain aggregate sizes. Aggregates on the order of metres can be easily eroded into centimetre sized aggregates, particularly with large ejected grains. With $\alpha = 5\xtenpow{-3}$, the results are shown in Fig.~\ref{Fig:Erosion_growth-3}, for grains subject to growth and erosion only.
In the left panel, the grains have an initial size of 0.2 $\mu$m and grow until they begin to drift towards the star as their Stokes number approaches $0.1$. The grains continue to grow while drifting until erosion occurs at a distance of $0.19~\au$. Erosion is extremely effective for ejected grains, with $s_\mathrm{ej}=1$ mm \citepalias{Rozner_2020_Erosiona}. The aggregates, with a maximum size of $4-6$ metres, are rapidly eroded to sizes of a few centimetres. Then, the grains reach an equilibrium between growth and erosion, and drift slowly from $0.19~\au$ inwards. For comparison, the green, orange and red lines represent the evolution of a grain considering growth and fragmentation, but no erosion. The fragmentation threshold is set to 20, 40 and 80 $\msec$ respectively. Even a high fragmentation threshold of 80 $\msec$ (a value completely unrealistic for most of the disc, resembling pure growth), the fragmentation threshold is reached before the erosion threshold. Therefore, fragmentation is always more effective in destroying dust grains when a realistic value is used.

\begin{figure*}
    \centering
    \includegraphics[width=0.48\textwidth,trim=.05cm .05cm .05cm .05cm,clip]{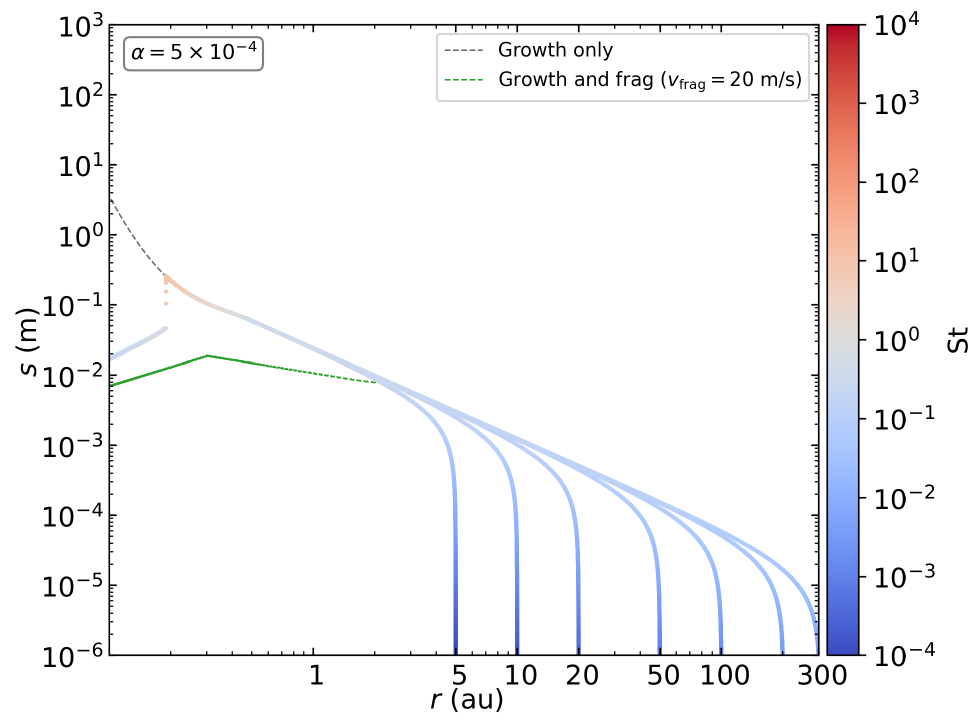}
    \includegraphics[width=0.48\textwidth,trim=.05cm .05cm .05cm .05cm,clip]{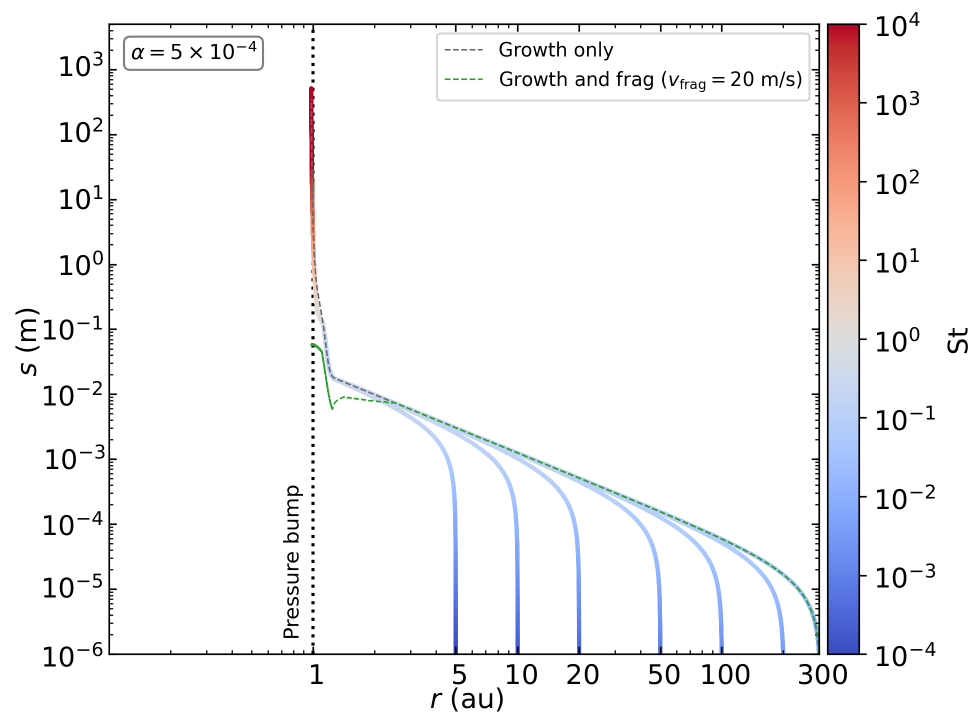}
    \caption{\textbf{Left}: Same as Fig.~\ref{Fig:Erosion_growth-3}, but with $\alpha = 5\xtenpow{-4}$ and only one fragmentation threshold.}
    \label{Fig:Erosion_growth-4}
\end{figure*}

We can also consider what would happen if the dust is trapped and can no longer drift to be accreted onto the star. The answer is shown in the right panel of Fig.~\ref{Fig:Erosion_growth-3}. Fragmentation is still effective in limiting the grain size. In the case without fragmentation, erosion is unable to balance out growth. There is a small difference around $s=0.1$ m due to erosion, which reduces the growth rate without completely countering it, affecting the final grain size somewhat. However, in this case, the final size is of little importance since the grains are trapped, making it possible for them to form larger objects of more than one kilometre at later stages.

The reason why erosion is almost absent within a dust trap is simple. Erosion depends on $\Delta v$, and in a dust trap, the dust-to-gas ratio ($\varepsilon=1$ in our case) is greater than the typical value of 1\%. Taking into account back-reaction, dust tends to make the gas orbit faster and vice versa, reducing $\Delta v$. Additionally, as the dust density $\rho_\mathrm{d}=\varepsilon\rho_\mathrm{g}$ increases, the growth rate is multiplied by 100 (Eq.~(\ref{Eq:Growth_dmdt_Stepinski_ter})). These two factors work against erosion and prevent it from occurring or at least having a significant impact.

We then perform the same simulations, but with $\alpha = 5\xtenpow{-4}$ shown in Fig.~\ref{Fig:Erosion_growth-4}. The results are qualitatively the same. 
In the left panel, the grains have the same initial size and grow until they begin to drift towards the star. The grains are eroded at the same distance of $0.19~\au$. Erosion is also effective, but with $\alpha = 5\xtenpow{-4}$, grain growth is less efficient because of lower relative velocities when the viscosity is lower. The aggregates, with a maximum size of 50 centimetres, are eroded to sizes of a few centimetres, reach the equilibrium and drift inwards. The green line represents the evolution when fragmentation is taken into account, with a threshold set to 20 $\msec$. Fragmentation is still more effective in destroying dust grains when such a realistic value is used. Higher thresholds with this $\alpha$ lead to an evolution resembling the pure growth case.
Nevertheless, erosion reduces the size of the grains by one order of magnitude, compared to roughly 3 when $\alpha = 5\xtenpow{-3}$.

Within the dust trap, shown on the right panel, erosion is almost absent.
The case with erosion does not differ from the case with pure growth, and only two scenarios are identified. If the threshold is low enough, grains are destroyed by fragmentation. Otherwise, dust grains can grow freely.

We also performed tests considering porosity \citep[see][for its implementation]{Michoulier_Gonzalez_Disruption}, and we observed no significant change in the appearance of erosion (not shown). We ran a simulation with $\alpha = 5\xtenpow{-5}$, but grain growth is so slow that dust barely reaches sizes of a few centimetres. 
Erosion still appears, but the dust size is only divided by a factor of two to three. In this case, the limiting factor to the grain size is the very low growth rate.

\begin{figure*}
    \centering
    \includegraphics[width=0.96\textwidth,clip]{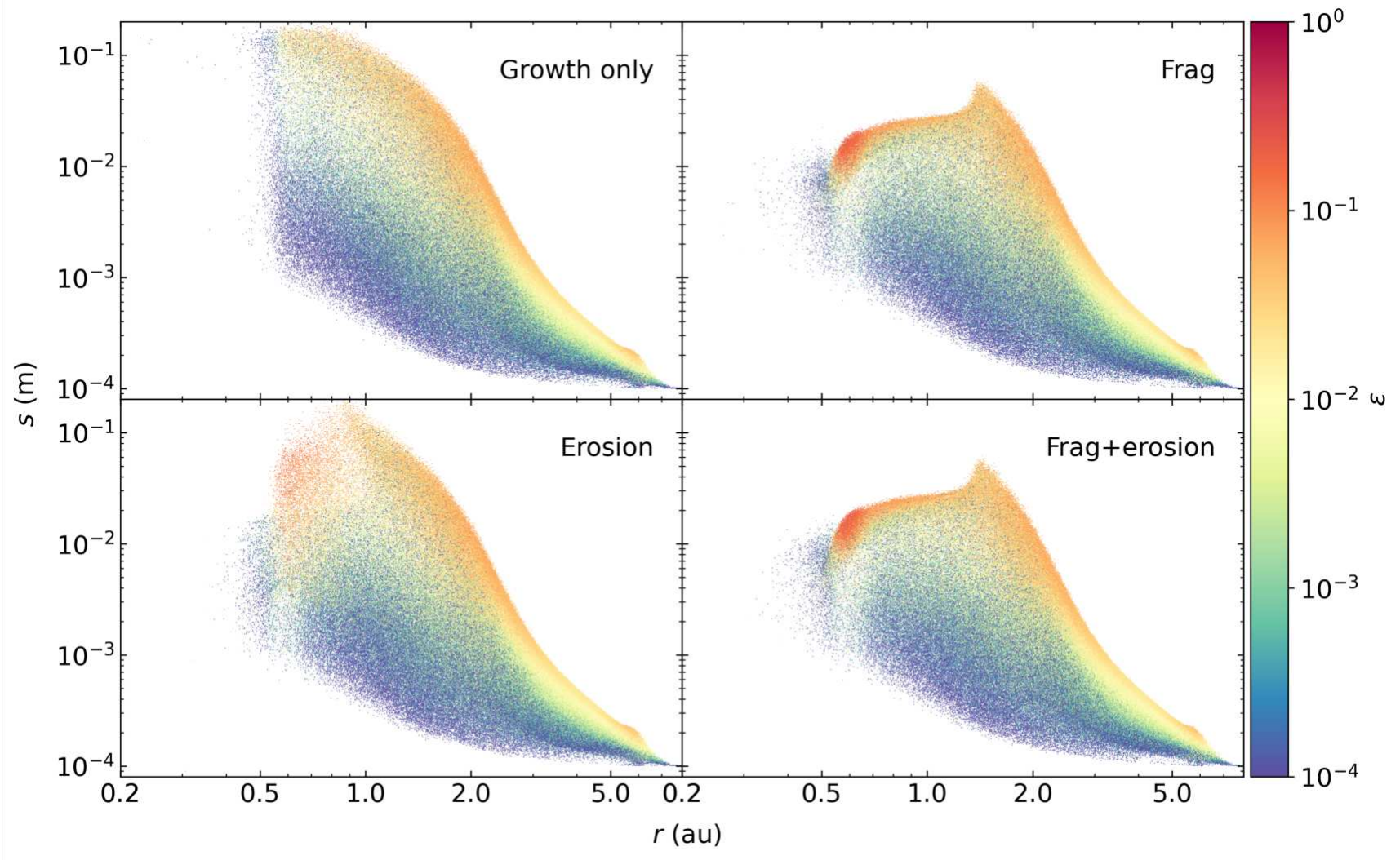}
    \caption{Comparison between simulations in the ($r$, $s$) plane at $t=100$ yr with growth only top left, growth and fragmentation top right with $v_\mathrm{frag,\:Si}=80~\msec$, growth and erosion bottom left and growth, fragmentation and erosion bottom right. The colour gives the dust-to-gas ratio.}
    \label{Fig:Erosion_growth-phantom}
\end{figure*}

\subsection{3D study}\label{Ssc:3D_Study}
We want to check now if we obtain the same behaviour with 3D simulations. We choose a high velocity threshold of $v_\mathrm{frag}=80~\msec$, for which fragmentation is hardly restrictive and our 1D simulations showed that it still dominates over erosion. Figure~\ref{Fig:Erosion_growth-phantom} shows the radial grain size distribution coloured with the dust-to-gas ratio $\varepsilon$, for four simulations at time $t=100$~yr. The top left panel corresponds to growth only, top right to growth and fragmentation, bottom left to growth and erosion, and bottom right to growth, fragmentation and erosion. We see that the simulation with growth is able to form large grains up to the decimetre between $0.5$ and $2~\au$. Simulations with fragmentation are identical when erosion is turned on or off (right panels). This means fragmentation is the dominant mechanism to destroy grains. When comparing the simulation with growth only and the one with erosion, the impact of erosion appears only when sizes are larger than $3$ cm, interior to $1~\au$. This is qualitatively similar to our 1D simulations, however here grains start to be eroded at larger distances ($\sim0.9$~$\au$). This is due to larger volume densities $\rho_\mathrm{g}$ in the midplane of the 3D disc compared to the 1D case. Moreover, fragmentation destroys grains at roughly the same size and at the same distance both in 1D and 3D.
Aggregates of decimetre size are still present, as in the growth-only simulation.
For the simulations with fragmentation, the peak in the size profile corresponds to the distance where $v_\mathrm{rel} = v_\mathrm{frag}$, and the plateau inwards shows an equilibrium between growth and fragmentation \citep{vericel_dust_2021}.

The simulation with growth only is the one where the inner region is the less dust-enriched, with $\varepsilon < \tenpow{-1}$. One should note that dust has been lost to the star due to radial drift in the very inner region. On the contrary, the ones with fragmentation have $\varepsilon \sim 2\xtenpow{-1}$ at 0.6 $\au$. Fragmentation helps grains to stay in the disc because their sizes become smaller the closer they are to the star, which means they will be more coupled to the gas and less prone to drift. With erosion only, the region between 0.6 and 1 $\au$ has $\varepsilon \sim \tenpow{-1}$. Erosion can thus help the inner region to increase the local dust-to-gas ratio, but it is not as efficient as fragmentation. Outside 2 $\au$, the size profiles of all simulations are very similar, as neither fragmentation nor erosion affect dust growth.

These simulations support the results from our 1D simulations. Moreover, they show that when accounting for fragmentation, erosion is not present even with $v_\mathrm{frag}=80~\msec$, which means it can be ignored during dust evolution and growth of aggregates.


\section{Discussion}\label{Sc:Discussion}
\subsection{Caveats}
The mains limitations arise from our model of dust growth, as we use a 3D code that eliminates most of the approximations done with the 1D Code \textsc{Pamdeas}. However, since we use the mono-disperse approximation for dust growth, we do not track all the smaller grains ejected from larger bodies due to erosion. This is not very important as we want to track the evolution of the largest grains, since fragmentation always appears before erosion. In addition, the mono-disperse approximation only considers collisions of equal-size grains. Unequal-size grains have relative settling and drifting velocities, acting as additional sources of collisions. The SPH formalism naturally produces a spread in $\Delta v$, resulting in a spread in $v_\mathrm{rel}$ as well, similar to the velocity distributions in \citet{windmark_velocity_distribution_2012,garaud_2013}. Nevertheless, our approximation can not account for high-mass-ratio collisions, which have been found to result in net growth even at $v_\mathrm{rel}$ as high as 70 or 80~m\,s$^{-1}$ \citep{teiser_2009,kothe_2010,windmark_sweep-up_2012,wada_2013,meisner_2013}.

Aeolian erosion is a continuous process and one may wonder whether the discrete nature of particle collision would hinder the comparison between both processes. This is not a problem in numerical simulations, where time is incremented by a finite quantity, the timestep. Even when the time between collisions is long, it is still much shorter than a single timestep \citep[e.g.][]{garcia_evolution_2020}. Both processes are thus taken into account simultaneously in our simulations.

Another issue is the fact that we do not take into account porosity in the model. This could play a role, as porous grains tend to form larger aggregates while being less sensitive to fragmentation. A model of erosion taking into account porosity would be more precise in capturing dust evolution. But we do not think this will change the results significantly because fragmentation will remain the limiting factor.

Lastly, in the erosion model, $\beta_\mathrm{eros}$ is still not very well constrained for different kinds of material and more experimental measurements would be needed, although progress has been made in recent work \citep{demirci_2020,schoenau_2023}. The model also assumes that all ejected grains have the same size, while in reality, all sizes should be considered. 

\subsection{Importance of erosion}
We see in this paper that fragmentation is always the first process to destroy grains before erosion for $v_\mathrm{frag} = 20~\msec$. Smaller fragmentation thresholds like 10 or 15 $\msec$ only strengthen fragmentation, thus decreasing the effect of erosion, while larger fragmentation threshold can be considered less realistic, both with $\alpha=5\xtenpow{-3}$ and $\alpha=5\xtenpow{-4}$. Moreover, we use the size $s_\mathrm{ej} = 1$ mm adopted by \citetalias{Rozner_2020_Erosiona} and \citetalias{Grishin_2020_Erosionb}, which makes erosion more effective. Therefore, we can safely say that erosion can be completely neglected in dust models when $\alpha=5\xtenpow{-3}$. For $\alpha=5\xtenpow{-4}$, the only fragmentation considered is $v_\mathrm{frag} = 20~\msec$, because higher $v_\mathrm{frag}$ are extremely similar to pure growth. The difference of sizes just before and after erosion in this case is also smaller, by an order of magnitude, reducing the impact of erosion on dust evolution. 
Finally, adding a pressure bump at $1~\au$ did not help erosion to appear, despite the fact that grains grow up to one kilometre. 
Erosion can therefore be neglected in all cases when fragmentation is considered. For a very low-viscosity disc, erosion could still be ignored, as grains growth would be very slow and large sizes wouldn't be reached.
However, it is worth mentioning that if larger boulders or planetesimals are formed or captured, erosion would still be important in grinding them down in the inner disc region, as discussed by \citet{Grishin_2019,Rozner_2020_Erosiona}.

\section{Summary and conclusion}
\label{Sc:Conclusions}

In this paper, we discuss the importance of erosion in the evolution of dust in protoplanetary disc.
We first present the way we modelled radial drift in a 1D code to capture accurately the effect even if the disc is stationary and the gas is not evolving. We then present how erosion has been treated based on \citetalias{Rozner_2020_Erosiona} and \citetalias{Grishin_2020_Erosionb}, giving the two important equation to implement in the codes. We also present the model to take into account both growth and fragmentation from \citet{stepinski_global_1997}, \citet{laibe_growth_2008}, \cite{garcia_evolution_2018}, \cite{vericel_dust_2021}, \cite{Michoulier_Gonzalez_Disruption}. We perform some tests to be sure \textsc{Pamdeas} reproduces some of the key results presented in \citetalias{Rozner_2020_Erosiona} and \citetalias{Grishin_2020_Erosionb}. We then perform simulations with the same model implemented in both the \textsc{Pamdeas} and \textsc{Phantom} codes.
We show that erosion is negligible when fragmentation is taken into account with realistic fragmentation thresholds. Both codes give the same results. When considering a pressure bump, erosion is still not efficient at destroying grains.
We then discuss the main caveats of this study and a discussion of the insignificance of erosion.
To conclude, erosion can be neglected in models of dust evolution accounting for fragmentation, bouncing, or any other mechanism that limits the dust sizes to a couple of centimetres in the very inner region of a protoplanetary disc. Usually, such short distances of a fraction of $\au$ are not considered when doing simulations and erosion can be safely neglected from a dust evolution perspective.

\begin{acknowledgements}
We thank Daniel J. Price for useful discussion and advice and the anonymous referee for their suggestions.
The authors acknowledge funding from ANR (Agence Nationale de la Recherche) of France under contract number ANR-16-CE31-0013 (Planet-Forming-Disks) and thank the LABEX Lyon Institute of Origins (ANR-10-LABX-0066) for its financial support within the Plan France 2030 of the French government operated by the ANR. This research was partially supported by the Programme National de Physique Stellaire and the Programme National de Planétologie of CNRS (Centre National de la Recherche Scientifique)/INSU (Institut National des Sciences de l’Univers), France. We gratefully acknowledge support from the PSMN (Pôle Scientifique de Modélisation Numérique) of the ENS de Lyon for the computing resources. This project has received funding from the European Union's Horizon 2020 research and innovation programme under the Marie Sk\l{}odowska-Curie grant agreements No 210021 and No 823823 (DUSTBUSTERS). Figures were made with the Python library \texttt{matplotlib} \citep{Hunter:2007}.
\end{acknowledgements}

%
%

\bibliographystyle{aa} 
\bibliography{erosion} 

\end{document}